\documentclass[reprint,amsmath,amssymb,aps]{revtex4-1}

\usepackage{epstopdf}
\usepackage{graphicx}
\usepackage{siunitx}

\begin{document}
\title{Electromagnetically induced transparency in a V-system with $^{87}$Rb vapour in the hyperfine Paschen--Back regime}

\author{Clare R Higgins and Ifan G Hughes}

\affiliation{Department of Physics, Durham University, South Road, Durham, DH1 3LE, UK}

\affiliation{Corresponding author: clare.r.higgins@durham.ac.uk}

\begin{abstract}
We observe EIT in a V-system in a thermal rubidium--87 vapour in the hyperfine Paschen--Back regime, realised with a  0.6\,T axial magnetic field. In this regime energy levels are no longer degenerate and EIT features from different initial states are distinct, which we show produces a much cleaner feature than without a magnetic field.  We compare our results to a model using the time-dependent Lindblad master equation, and having averaged over a distribution of interaction times, see good qualitative agreement for a range of pump Rabi frequencies. Excited state decay into both ground states is shown to play a prominent role in the generation of the transparency feature, which  arises mainly due to transfer of population into the ground state not coupled by the probe beam. We use the model to investigate the importance of coherence in this feature, showing that its contribution is more significant at smaller pump Rabi frequencies. 
\end{abstract}

\maketitle

\section{Introduction}
Electromagnetically induced transparency (EIT) is an optical phenomenon involving three quantum states coupled by two optical fields (laser beams). In an absorbing medium, a transparency window in the transmission of a weak probe beam on one transition is induced by the presence of a strong pump beam on another transition \cite{ARIMONDO1996257}. Throughout the text we refer to these beams as `pump' and `probe'. EIT has been widely studied and has potential applications in precision magnetometers \cite{Fleischhauer:magnetometry,Budker2007:magnetometry, Yudin:magnetometry}, slow light generation \cite{Hau:slowlight, Das_2018}, quantum information \cite{Beausoleil:quantuminformation, Hammerer:quantuminformation}, and atomic clocks \cite{Santra:atomicclock}. There are three possible configurations of EIT: V; lambda; and ladder~\cite{FleischhauerReview}. V-EIT is the least studied of these because there is no stable dark state \cite{KHAN20164100}, as both of the singly coupled states are excited states and can decay to the ground state. Nevertheless, V-EIT has been extensively studied \cite{DEY20152711, Hazra_2020, Das_2018, Hoshina:14, PhysRevA.57.1323, PhysRevA.83.063419, PhysRevA.59.4675, PhysRevA.87.043813, McGloin_2003, PhysRevA.71.053806, ZHAO2002341, PhysRevA.92.063810, CHA2014175, Das:21, VDOVIC2007407}, and provides an interesting testing ground for ascertaining the relative importance of coherent and incoherent mechanisms in the generation of the transparency window~\cite{Kang:17,   PhysRevA.52.2302, PhysRevA.83.063419}.

One of the main obstacles to overcome in modelling and understanding V-EIT in thermal vapours is the complexity introduced by the overlapping spectral lines, as a consequence of the degeneracies of the magnetic sub-levels and the excited-state hyperfine splitting being less than the Doppler width of the probed transition.
To circumvent these difficulties, we use the  hyperfine Pashen-Back regime \cite{PhysRevA.84.063410, Weller_2012, Zentile_2014, Ponciano_Ojeda_2020, Sargsyan:14, Sargsyan:17a,  PhysRevA.95.061804, doi:10.1063/1.4993760} where the energy levels are non-degenerate. A 0.6\,T magnetic field used with $^{87}$Rb vapour on the D1 and D2 lines leads to isolated transitions separated by more than their Doppler width. Previous work has shown that operating in this regime allows simplified energy-level schemes and theoretical models, leading to good agreement between theory and experiment~\cite{WhitingEIT, WhitingEIA, Whiting:FWM,WhitingSinglephotons}.

\section{Theory}
\label{sec:theory}

\begin{figure}[htb]
\centering
\includegraphics[width=0.9\linewidth]{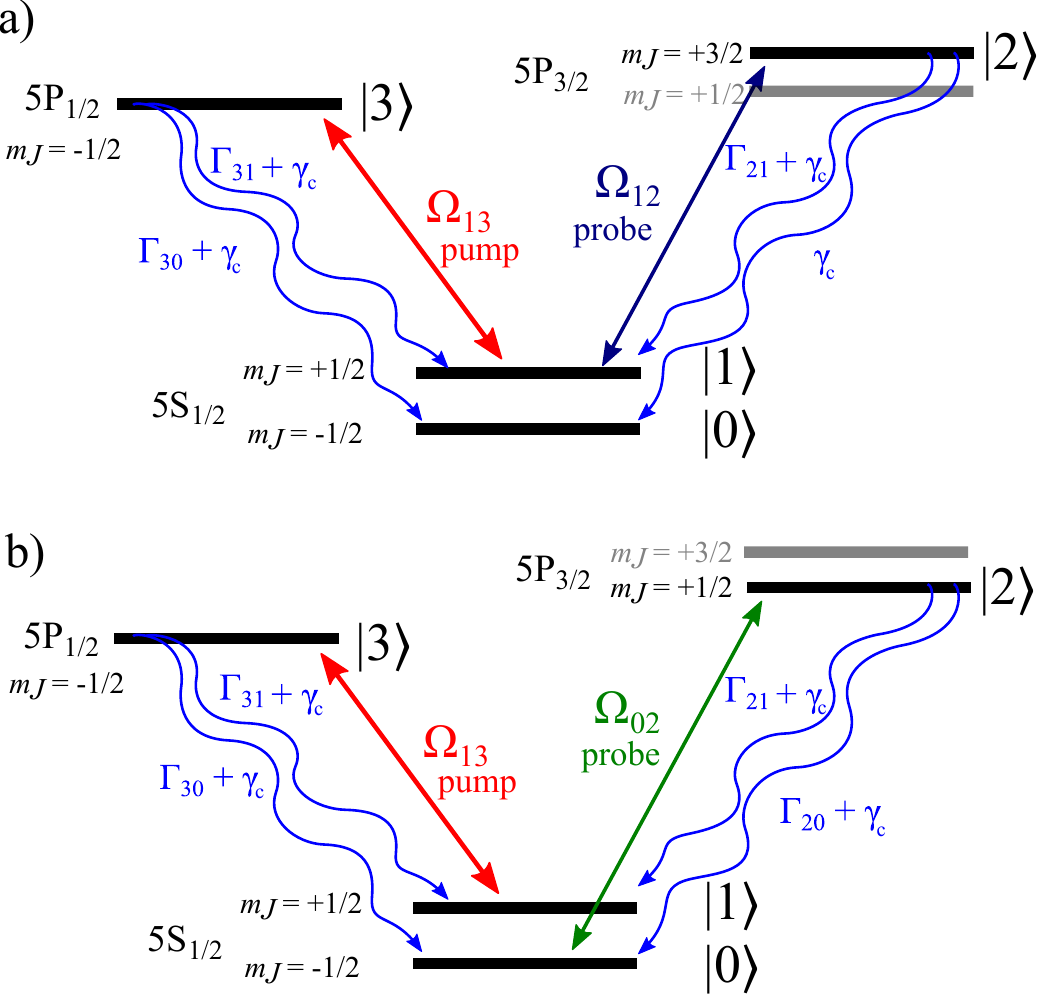}

\caption{The energy levels involved in our system. a) shows the EIT configuration, in which the 780\,nm probe beam is coupling $m_J$ = +1/2 to $m_J$ = +3/2. In b) the probe beam is instead tuned to couple the $m_J$ = -1/2 to $m_J$ = +1/2 transition, which excites out of $|0\rangle$, the non-pump-coupled ground state. This configuration does not produce EIT, but demonstrates that population moves from $|1\rangle$ to $|0\rangle$. These two probe positions produce the set~1 and set~2 of peaks in Fig\,\ref{fig:full-scan}, respectively. $\Omega_{ab}$ are the driving Rabi frequencies, between initial state $a$ and final state $b$. The decays between states have two contributions: $\Gamma_{ab}$, the natural linewidth, and $\gamma_c$, the collisional decay to each ground state. The probe(pump) is left(right)-hand circularly polarised and couples  $\sigma_+$($\sigma_-$) transitions~\cite{f2f}.}

\label{fig:energy-levels}
\end{figure}

Our V-EIT system, realised in the hyperfine Paschen-Back regime, is shown in Fig. \ref{fig:energy-levels}, part a). The levels we use, marked $|1\rangle$, $|2\rangle$ and $|3\rangle$, do not form a closed system. We use `closed system' to mean the atoms do not decay to any states outside of the three EIT levels, and `open system' when decay to other, non laser-coupled states, is possible.  The pumped transition -- from $|1\rangle$ to $|3\rangle$ -- is an open transition so $|3\rangle$ can decay to the other, uncoupled, ground state. This adds a fourth level into the system, which we label $|0\rangle$. The pump causes population transfer from $|1\rangle$ to $|0\rangle$, resulting in reduced absorption of the probe which couples $|1\rangle$ and $|2\rangle$. The driving Rabi frequencies are labelled $\Omega_{ab}$, where $a$ and $b$ represent the initial and final states respectively. The decays between states have two contributions: the natural linewidth, $\Gamma_{ab}$, and a collisional decay to each ground state, $\gamma_c$. The second is present even where dipole-allowed transitions are forbidden, and the total collisional decay from an excited state has been experimentally determined in this vapour cell as $2\gamma_c /2\pi$ = \SI{7}{\mega\Hz} \cite{WhitingEIT}. The natural linewidths (linear) of states $|2\rangle$ and $|3\rangle$ are \SI{6.0}{\mega\Hz} and \SI{5.7}{\mega\Hz}, respectively. These are split along the two decay paths according to the branching ratios calculated using Wigner 3-$j$ symbols. Part b) shows the state configuration when the probe is instead tuned to the transition between $|0\rangle$ and $|2\rangle$. This is not an EIT setup, but allows us to see the enhanced absorption caused by the extra population in $|0\rangle$.

\section{Experimental Details}

\begin{figure}[htb]
\centering
\includegraphics[width=\linewidth]{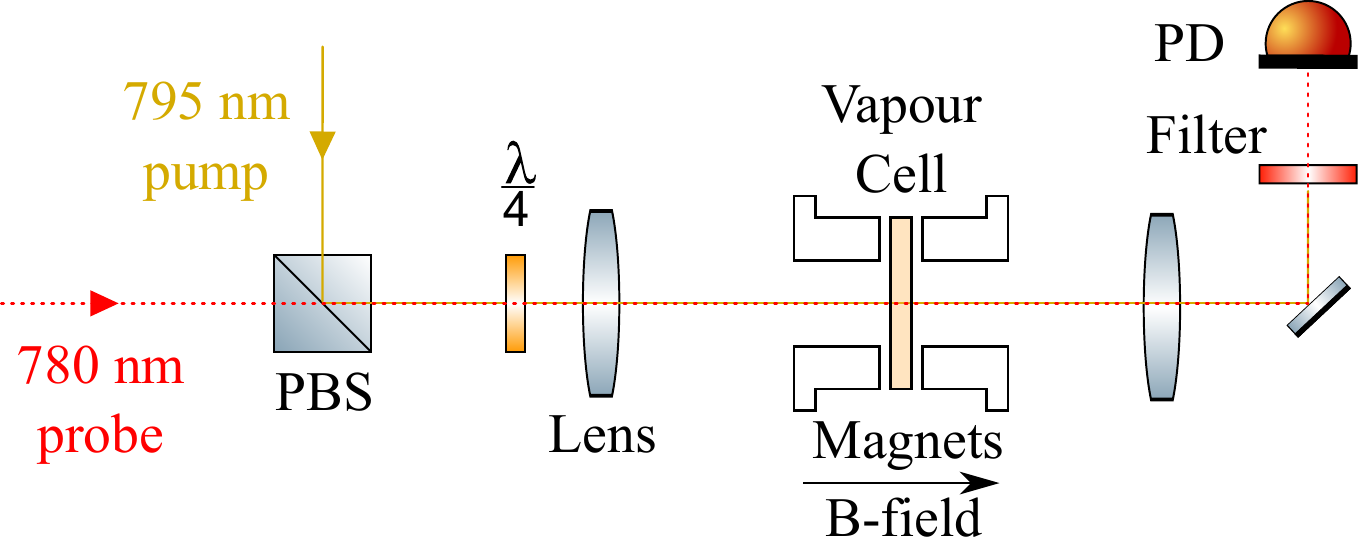}

\caption{Experimental setup. Orthogonally linearly polarised 795\,nm pump and 780\,nm probe beams are combined on a polarising beam splitter (PBS) cube, and passed through a quarter waveplate converting them to right- and left-handed circularly polarised light, respectively. The beams are focused through a 2\,mm vapour cell in a longitudinal 0.6\,T magnetic field, to an average beam waist of \SI[separate-uncertainty]{83\pm5}{\micro\meter}.The light transmitted through the cell passes through an interference filter to remove pump light, and is recorded on a photodiode (PD).}
\label{fig:setup}
\end{figure}

The experimental setup is shown in Fig.\,\ref{fig:setup}. We use a 2\,mm long 98$\%$ $^{87}$Rb vapour cell in a magnetic field,  parallel to the laser propagation direction, of 0.6\,T, produced by two cylindrical `top hat' magnets. The orthogonally linearly polarised 795\,nm and 780\,nm beams are combined on a polarising beam splitter. A quarter waveplate transforms the polarisation to left-hand circular and right-hand circular respectively. A  lens of focal length 200\,mm focusses the beams to waists of \SI[separate-uncertainty]{100\pm5}{\micro\meter} $\times$ \SI[separate-uncertainty]{78\pm5}{\micro\meter} (780\,nm) and \SI[separate-uncertainty]{65\pm5}{\micro\meter} $\times$ \SI[separate-uncertainty]{90\pm5}{\micro\meter} (795\,nm) inside the cell.  We aim to overlap the beams as completely as possible inside the cell by optimising the EIT feature, however due to the slight shape difference a perfect overlap is not possible. After the cell an interference filter removes pump light, and the probe transmission spectrum is measured on a photodiode. We have a strong, resonant 795\,nm pump, and a weak 780\,nm scanning probe. We use a vapour temperature of \SI{80}{\celsius}; at lower temperatures the signals are smaller, and at higher temperatures the absorption saturates and the features are distorted.

\section{Experimental Results}
\begin{figure}[htb]
\centering
\includegraphics[width=\linewidth]{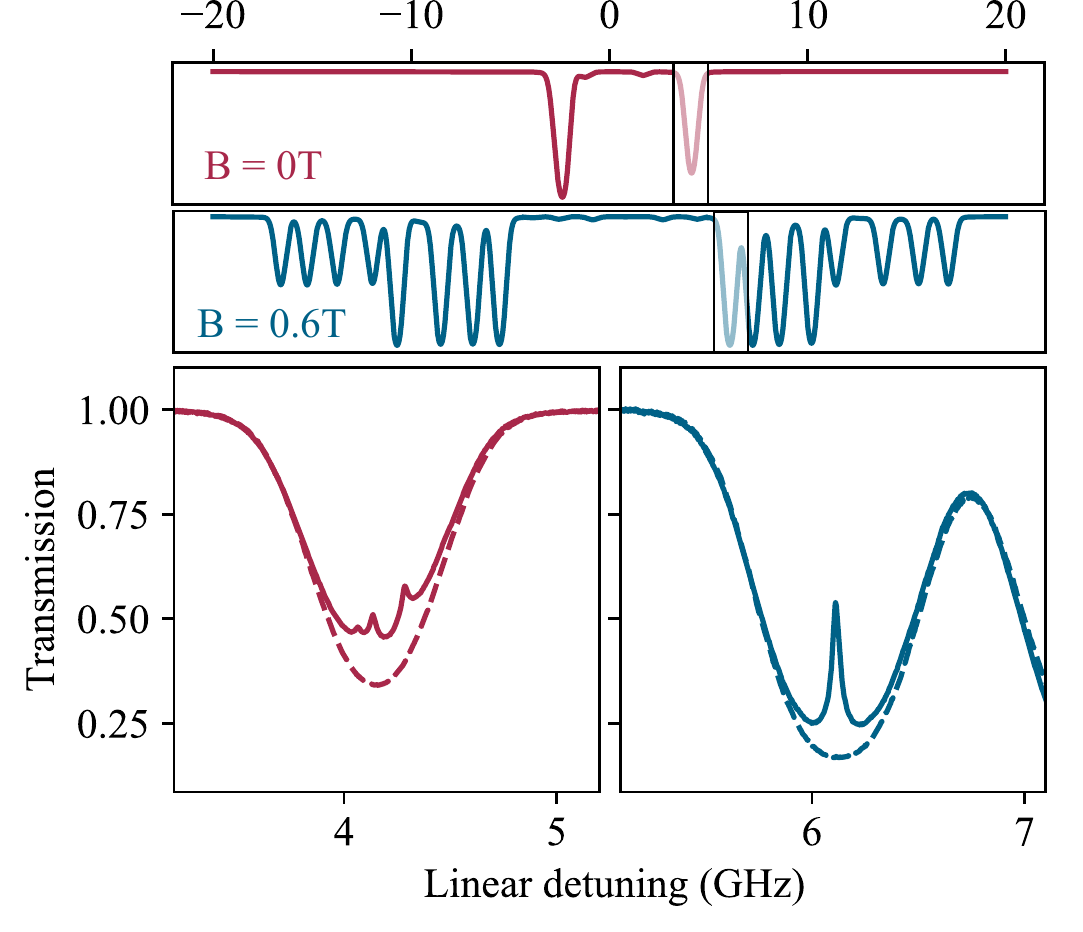}

\caption{Upper panels: Theoretical scans over D2 features without magnetic field (top, red), and with 0.6\,T magnetic field (second panel, blue). Shaded rectangles show where the experimental spectra in the lower panels fit in the spectra. In the lower panels dotted lines are probe beam only, solid lines are when the pump beam is introduced. Left: Experimental V-EIT feature with no magnetic field. Many hyperfine sublevels contribute producing a messy feature. Right: Experimental feature in 0.6\,T magnetic field. Energy levels are separated by more than the Doppler width so a single clean feature is seen.}

\label{fig:no-mag}
\end{figure}

\begin{figure}[htb]
\centering
\includegraphics[width=0.8\linewidth]{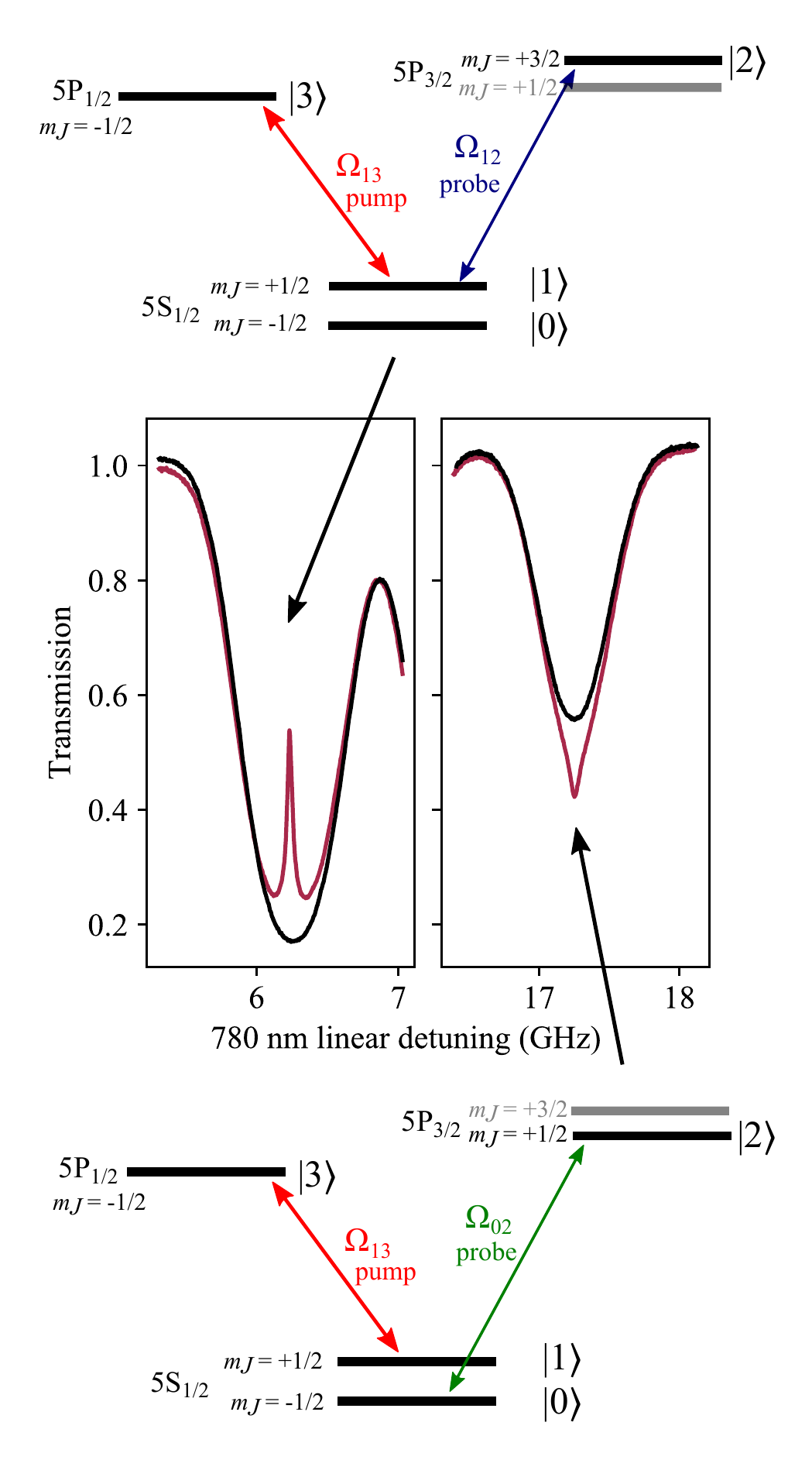}

\caption{Figure shows a \SI{0.1}{\micro\watt}, 780\,nm probe beam scan over the two $m_I = 3/2 $ D2 transition lines at 0.6\,T. The black trace is probe only, while the red traces shows the effect of introducing a \SI{20}{\micro\watt} pump beam on the transition $|1\rangle \leftrightarrow |3\rangle$, as shown in the energy level diagrams. The energy level diagrams show which states the pump beam is coupling for the two features in the scan. Where the probe couples out of $|1\rangle$, there is a transparency feature, and where it couples out of $|0\rangle$ there is an enhanced absorption feature. } 
\label{fig:full-scan-1-set}
\end{figure}

\begin{figure*}[htb]
\centering
\includegraphics[width=\linewidth]{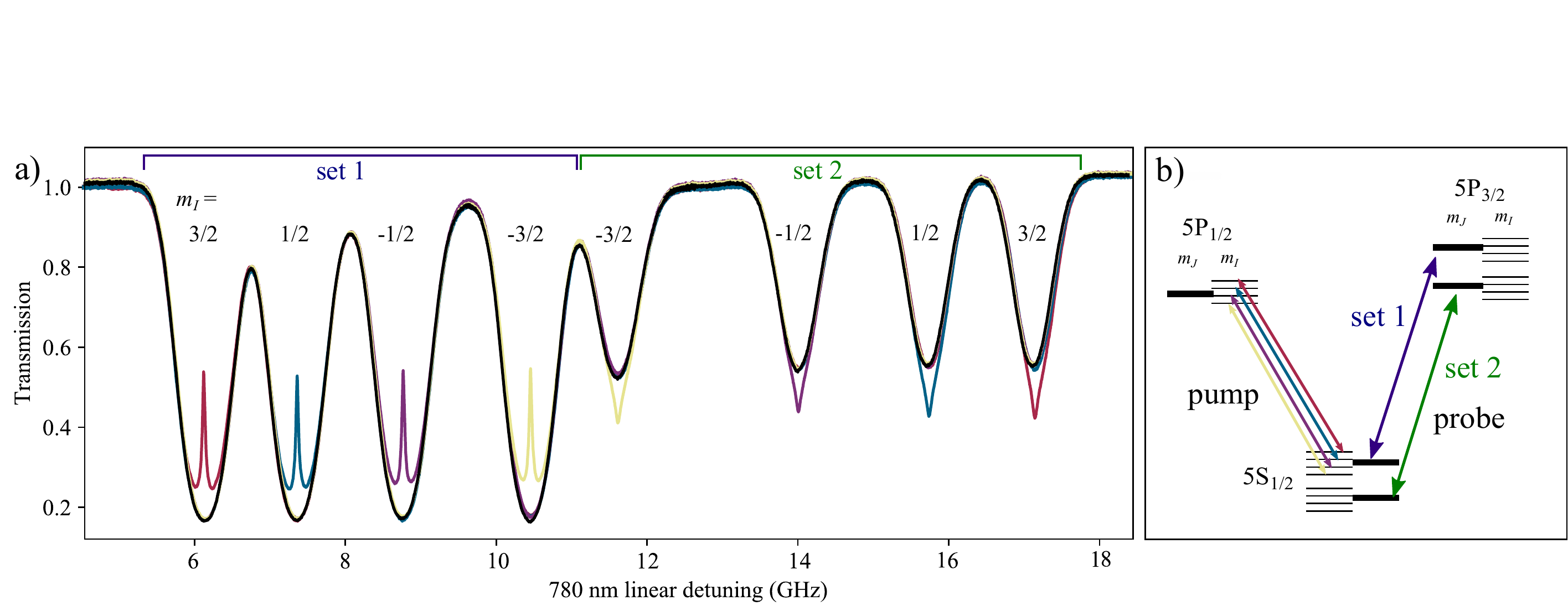}

\caption{a) shows a \SI{0.1}{\micro\watt}, 780\,nm probe beam scan over the D2 transition lines at 0.6\,T. The black trace is probe only, while the four coloured traces show the effect of introducing a \SI{20}{\micro\watt} pump beam on the correspondingly coloured transition shown in b). Each of the four pump transitions has a different $m_I$ value. Introducing a particular $m_I$ pump transition induces a transparency in the peak in set~1, and an enhanced absorption feature in the corresponding peak in set~2, which have the same $m_I$.}
\label{fig:full-scan}
\end{figure*}

\begin{figure}[htb]
\centering
\includegraphics[width=\linewidth]{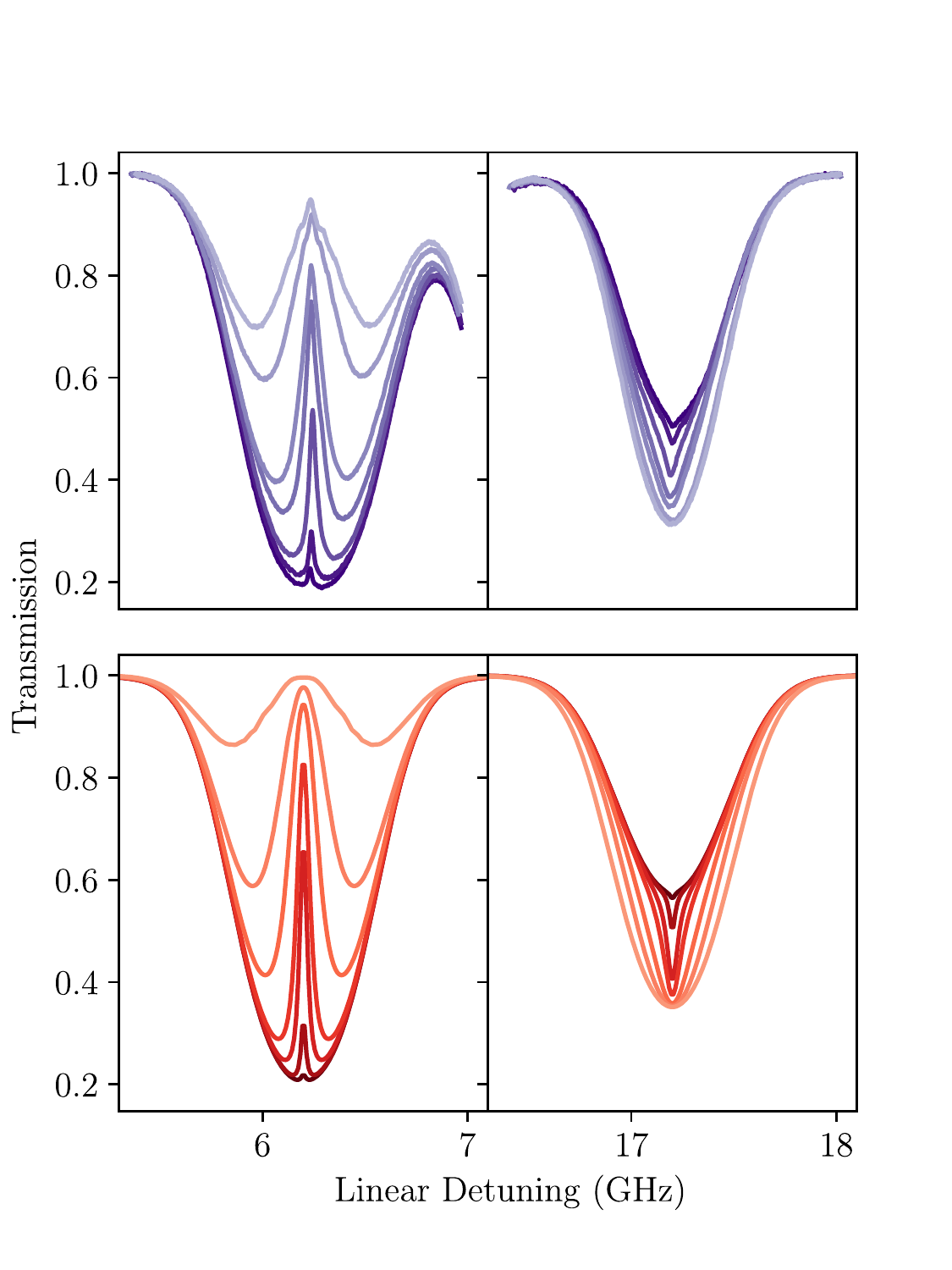}

\caption{The effect of changing the 795\,nm pump power on the induced transparency and enhanced absorption features on the $m_I = +3/2$ transitions of the D2 (780\,nm, \SI{0.1}{\micro\watt}) spectrum. Upper: Experimental transmission spectra with changing pump powers, with values of in \SI{}{\micro\watt} of 1 (dark), \SI{5}{}, \SI{10}{}, \SI{50}{}, \SI{100}{}, \SI{500}{}, \SI{1000}{} (light). These correspond to Rabi frequencies in the range \SIrange{2}{100}{\mega\Hz}. The EIT features shown in the top left panel have FWHM ranging from \SI[separate-uncertainty]{21\pm3}{\mega\Hz} (lowest pump power) to \SI[separate-uncertainty]{247\pm5}{\mega\Hz} (highest pump power). Lower: Modelled transmission spectra with pump Rabi frequencies in \SI{}{\mega\Hz} of 1 (dark), \SI{3}{}, \SI{10}{}, \SI{20}{}, \SI{50}{}, \SI{100}{}, \SI{300}{} (light). The range of Rabi frequencies was chosen to straddle the range of features seen in the experimental data; they are not calculated equivalents. }
\label{fig:change-power}
\end{figure}
Fig.\,\ref{fig:no-mag} shows the advantage gained by using the hyperfine Paschen-Back regime. The top panels show theoretical D2 line spectra without a magnetic field (red) and with a 0.6\,T field (blue). The two peaks used in the lower panels are shown highlighted. In the lower panels dotted lines are probe beam only; solid lines are when the pump beam is introduced. The probe only features have a Voigt profile with FWHM of $\sim$ \SI{550}{MHz} at \SI{80}{\celsius}. The profile is dominated by its Gaussian component, which is due to the Doppler effect; atoms at finite temperature travel at a range of velocities which each absorb at a frequency displaced from resonance, given by $\omega = \omega_0 + k v$. Here $\omega$ is angular frequency, $\omega_0$ is resonance angular frequency, $k$ is wavenumber and $v$ is the velocity component along the direction of propagation of the laser beams. The left panel shows experimental EIT features with no magnetic field, which shows contributions from several transitions. The right shows the feature in a 0.6\,T field, where one clean feature is visible. Both features are produced in the same cell, with the same laser powers. 

Fig.\,\ref{fig:full-scan-1-set} shows a scan over the two $m_I = 3/2 $ D2 $\sigma_+$ absorption lines in a 0.6\,T magnetic field. The black trace is a probe only scan, and the red traces shows the effect of adding in a \SI{20}{\micro\watt} pump beam. Here, and throughout, we use a probe power of \SI{0.1}{\micro\watt}.  All the optical power values reported throughout this work are measured before the vapour cell, and have an error of $\pm5\%$. We see that two different features appear; on the $m_J = -1/2 \rightarrow m_J = +1/2$ peak (left) we see a narrow transmission feature, characteristic of EIT. The states coupled at this point in the scan are shown in the diagram above. Notably the probe is coupling out of $|1\rangle$, the upper ground state. On the $m_J = -1/2 \rightarrow m_J = +1/2$ peak (right) there is an enhanced absorption feature. In this case, as shown in the lower diagram, the probe couples out of $|0\rangle$, the lower ground state. This state is populated by spontaneous decay from $|3\rangle$, which is itself populated by the strong pump beam. 

Following on from Fig.\,\ref{fig:full-scan-1-set}, Fig.\,\ref{fig:full-scan} shows the effect of tuning the pump beam to different $m_I$ transitions. The black trace in a) shows a scan of the 780\,nm probe over the D2 absorption lines at \SI{0.6}{\tesla}, with no pump. 
At 0.6\,T,  $m_I$ and $m_J$ are good quantum numbers. For $^{87}$Rb, $I=3/2$, therefore there are four possible values for $m_I$. The spectrum shows two sets of four transitions; in set~1(set~2) all four transitions are between states with initial $m_J = +1/2(-1/2)$ and final $m_J = +3/2(+1/2)$. Inside each set, each transition has a different $m_I$ value, as labelled in the figure. The four coloured traces show the probe transmission when the pump is tuned to the correspondingly coloured transition in b). It is evident that when the pump is coupled to a particular $m_I$ level in the upper ground state, $|1\rangle$, there is a transmission window in the probe absorption peak coupling out of that level (set~1 transitions). There is also a corresponding enhanced-absorption feature when the probe instead couples out of the lower ground state with the same $m_I$ value, $|0\rangle$ (set~2 transitions). The EIT features shown in Fig.\,\ref{fig:full-scan} have a FWHM of \SI[separate-uncertainty]{39\pm3}{\mega\Hz}.

\section{Model}
Atomic systems can be modelled using the Lindblad-Master equation \cite{FleischhauerReview},
\begin{equation}
    \frac{\text{d}\rho}{\text{d}t} = - \frac{i}{\hbar}[H, \rho] + L,
\end{equation}
which describes the evolution of the density matrix, $\rho$,
\begin{equation}
    \rho = 
    \begin{pmatrix}
        \rho_{00} & \rho_{10} & \rho_{20} & \rho_{30} \\
        \rho_{01} & \rho_{11} & \rho_{21} & \rho_{31} \\
        \rho_{02} & \rho_{12} & \rho_{22} & \rho_{32}  \\
        \rho_{03} & \rho_{13} & \rho_{23} & \rho_{33}  
    \end{pmatrix},
\end{equation}
of the system. The diagonal elements, $\rho_{aa}$, are the population in each state, and the off-diagonal elements, $\rho_{ab}$, are the coherences between states. The system Hamiltonian, $H$, in the rotating wave approximation, has state detunings, $\Delta_{ab}$, on the diagonals, and Rabi frequencies, $\Omega_{ab}$, coupling the states on the off-diagonals.
The Hamiltonian corresponding to the system in Fig.\,\ref{fig:energy-levels} a) is 
\begin{equation}
    H = \frac{\hbar}{2}
    \begin{pmatrix}
        0 & 0 & 0 & 0 \\
        0 & 0 & \Omega_{12} & \Omega_{13} \\
        0  & \Omega_{12} & -2\Delta_{12} & 0  \\
        0 & \Omega_{13} & 0 & -2\Delta_{13}
    \end{pmatrix},
\end{equation}
while the Hamiltonian for Fig.\,\ref{fig:energy-levels} b) is
\begin{equation}
    H = \frac{\hbar}{2}
    \begin{pmatrix}
        0 & 0 & \Omega_{02} & 0 \\
        0 & 0 & 0 & \Omega_{13} \\
        \Omega_{02}  & 0 & -2\Delta_{02} & 0  \\
        0 & \Omega_{13} & 0 & -2\Delta_{13}
    \end{pmatrix}.
\end{equation}

As we use a V system in a co-propagating geometry, we incorporate the Doppler effect into the model by setting $\Delta_{\text{pump}} \rightarrow \Delta_{\text{pump}} - k_{\text{pump}} v$ and $\Delta_{\text{probe}} \rightarrow \Delta_{\text{probe}} - k_{\text{probe}} v$. Prominent EIT features are observed with velocity groups where the residual two-photon doppler broadening $(k_{\text{pump}}-k_{\text{probe}})v < \Omega_{\text{pump}}$ \cite{PhysRevA.52.2302}. This geometry makes the system Doppler insensitive, because the two photon resonance condition is maintained for atoms of non-zero velocity. 

Decays between states are included in the Lindblad dissipator term, $L$, given by
\begin{equation}
    L = \sum_n \frac{1}{2} [2C_n\rho C_n^\dagger - (\rho C_n^\dagger C_n + C_n C_n^\dagger \rho)],
\end{equation}
which is a sum over all decay modes, $n$, where $C_n = \sqrt{\gamma_n}A_n$ are collapse operators and $A_n$ are operators which couple the environment to the system with rate $\gamma_n$. For our system this means $C_{ab}~=~\sqrt{\Gamma_{ab}+ \gamma_c} |b\rangle \langle a|$.

We solve the Lindblad master equation numerically for our four-level system. We model with a probe beam Rabi frequency of \SI{0.96}{\mega\Hz} on the $m_J~=~+~1/2~\rightarrow~m_J~=~+~3/2$, and, due to the differing dipole matrix elements of the transitions, a Rabi frequency of \SI{0.55}{\mega\Hz} on the $m_J~=~-1/2~\rightarrow~m_J~=~+1/2$ transitions. This puts us in the weak probe regime. We use a range of pump Rabi frequencies to produce a range of features which span those seen experimentally. This range is \SIrange{2}{100}{\mega\Hz}.
The pumping transition we use is open, as the excited state, $|3\rangle$, can decay to both $m_J$ ground states, $|0\rangle$ and $|1\rangle$, as depicted in Fig.~\ref{fig:energy-levels}a). The pump and probe only couple to $|1\rangle$ so in the steady-state solution all the population ends up in the uncoupled ground state, $|0\rangle$, resulting in no absorption. We therefore have to use the time-dependent solutions.  We have beams with an average $1/e^2 $ radius of \SI[separate-uncertainty]{83\pm5}{\micro\meter}, from which we calculate the in-beam time-of-flight distribution, using the transverse velocity distribution~\cite{PhysRevA.73.062509, doi:10.1143/JPSJ.78.084302, Sagle_1996}. For a given probe detuning the solutions are summed over all longitudinal velocity contributions~\cite{doi:10.1080/09500340.2017.1328749}.
The absorption profile is calculated from  the imaginary part of the relevant coherence, which is a suitably weighted average of the result at each time step, and we use the Elecsus code~\cite{Zentile2015b} to calculate the linestrengths.

\section{Comparison with experiment}
Fig.\,\ref{fig:change-power} shows the effect of changing pump power/Rabi frequency on the transmission and absorption features, with experimental results in the upper panels and  model predictions in the lower panels. Optical power is related to Rabi frequency by the area of the beam and the dipole matrix element of the transition. Here, as we are not plotting theory and experiment on the same axis we use optical power for experiment, and Rabi frequency for theory. We see good qualitative agreement, with both the narrow transmission and the extra absorption feature correctly predicted, though the features are slightly narrower in theory than in experiment. We attribute the small sub-features seen in the 0.5 and 1 mW spectra to back reflections, which become more significant at higher powers. 
\begin{figure}[hb]
\centering
\includegraphics[width=0.7\linewidth]{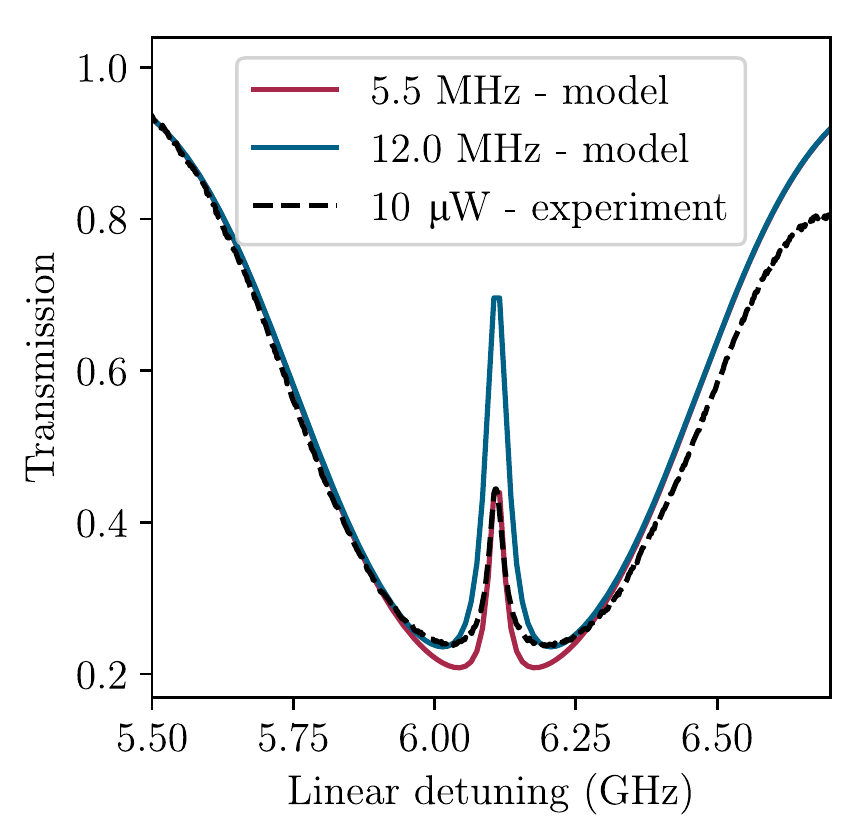}

\caption{A comparison of experimental and modelled results. Black dashed line is experimental with \SI{10}{\micro\watt} pump, which converts to an average linear Rabi frequency in the beam of 15 MHz. Two theory traces are plotted, chosen to fit the tip of the transparency feature (red) and the bottom of the absorption feature (blue). They have linear Rabi frequencies of \SI{5.5}{\mega\Hz} and \SI{12.0}{\mega\Hz} respectively, which are close to our experimental value. It is clear that for this model a pump Rabi-frequency cannot be chosen which fits well to all aspects of the feature; we must choose one or the other.}
\label{fig:numeric-comparison}
\end{figure}
Fig\,\ref{fig:numeric-comparison} shows the experimental trace for \SI{10}{\micro\watt} pump power, plotted with \SI{5.5}{\mega\Hz} and \SI{12.0}{\mega\Hz} model predictions. We see that using this model, we can choose to fit the peak of the absorption window, or the depth of the feature, but not both at once. In our model we assume that the beams have uniform intensity, whereas in reality they have a Gaussian profile; consequently atoms will experience a varying pump intensity as they traverse the beam. The intensity they see is also correlated to the time they spend in the beam. These factors are likely to change the shape of the spectra. For a numerical comparison we use the average Rabi frequency of the beam within the $1/e^2 $ waist, and an average power through the cell, taking into account absorption along its length and at cell windows. In this way, \SI{10}{\micro\watt} input power converts to an average Rabi frequency in the cell of \SI{15}{\mega\Hz}, which is close to the two model values. Both the issues mentioned above, and the fact that we don't fully take into account the incomplete spatial overlap of the beams could explain this discrepancy. A more detailed numerical model beyond the scope of this work is required to fully account for the shape of the EIT features. 
% \begin{figure}[hb]
% \centering
% \includegraphics[width=\linewidth]{figures/detuning-data-and-theory-grid-diff-colours-correctx.pdf}

% \caption{The effect of changing the pump field detuning on the \SI{0.1}{\micro\watt} 780\,nm probe transmission and enhanced absorption features, for $m_I = +3/2$ transitions. The top two panels show experimental results, with a pump power of \SI{20}{\micro\W}. Bottom panels show the model prediction, using a pump Rabi frequency of \SI{18}{\mega\Hz}. Dark to light traces are increasing pump frequency.}
% \label{fig:change-detuning}
% \end{figure}

% Fig\,\ref{fig:change-detuning} shows the effect of changing pump detuning whilst keeping pump power constant at \SI{20}{\micro\watt}, and scanning the probe. Top panels show experimental results and bottom panels show model prediction.  Changing the pump detuning selects atoms with a different longitudinal  velocity, and concomitantly changes the frequency of the transmission (left) and enhanced absorption (right) features in the probe scan. This behaviour is predicted by the model, as shown in the lower two panels. 

\section{How significant is the coherent effect?}
\begin{figure}[htb]
\centering
\includegraphics[width=0.8\linewidth]{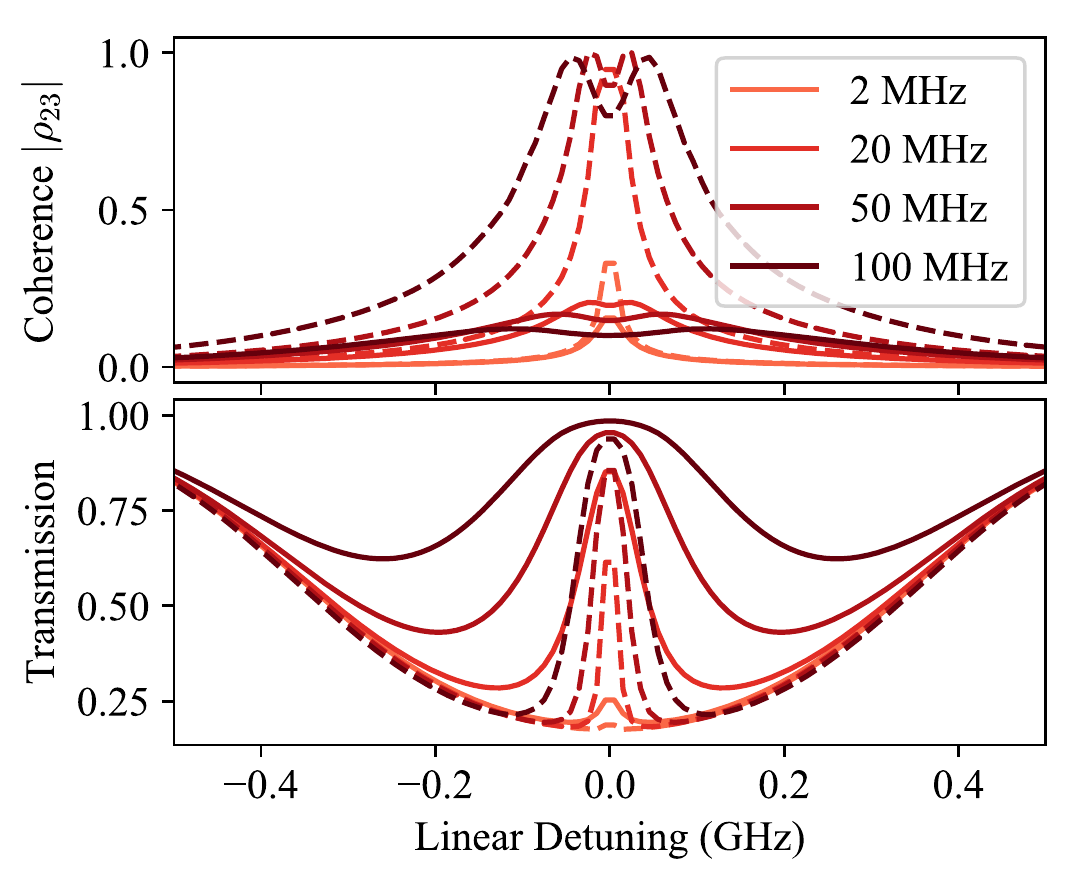}

\caption{ Plot showing the absolute value of the coherence between excited states $|\rho_{23}|$, as extracted from the model, and the corresponding probe transmission. On each plot we compare a closed system (dashed lines), and the open system of our experiment (solid lines), for linear pump Rabi frequencies as shown in the legend.}
\label{fig:coherences}
\end{figure}

A relevant question in three-level-systems is whether the spectral features are caused by coherent or incoherent effects \cite{PhysRevA.83.063419, KHAN20164100, Kang:17}. The presence of a prominent enhanced absorption feature on the transition out of the non-pump-coupled ground state, $|0\rangle$, is evidence that a significant part of the transmission feature does not arise from a coherent EIT effect, but instead from population transfer to a different (and uncoupled) ground state via velocity-selective optical pumping. However, the coherent process is still present, and we can use the model to see this. The density matrix element $\rho_{23}$ is the coherence between $|2\rangle$ and $|3\rangle$, the excited states of our system. Fig.\,\ref{fig:coherences} compares the transmission (lower panel) and corresponding coherence (upper) for a closed system -- meaning no decays into $|0\rangle$ -- (dotted lines) and our open system (solid lines). A range of pump Rabi frequencies are plotted and coloured according to the legend. We see that as pump Rabi frequency increases, the difference between the coherences in the closed and open systems increases, and that in our system, coherence increases as Rabi frequency increases up to a point (approx. 20 MHz), above which coherence decreases. We also note that for a given pump Rabi frequency the closed system coherence is greater than the open system coherence, while the open system transmission is greater than the closed system transmission. This shows that in our open system the coherence is a small, but present, cause of the feature and that as the pump Rabi frequency increases, its proportional contribution decreases. 

% \begin{figure}[htb]
% \centering
% \includegraphics[width=0.8\linewidth]{figures/mag_rho_23_cs_v1.pdf}

% \caption{Plot showing the coherence between excited states, as extracted from the model, and the corresponding probe transmission. On each plot we compare a closed system (dashed lines), and the open system of our experiment (solid lines), for linear pump Rabi frequencies as shown in the legend.}
% \label{fig:coherences}
% \end{figure}

% \begin{figure}[htbp]
% \centering
% \includegraphics[width=\linewidth]{figures/fig-4-data-and-theory-grid-v2.pdf}

% \caption{780 pump, 795 probe. Figure showing the effect of changing the 780\,nm pump power on the EIT and EIA features of the D1 (795\,nm) spectrum. Upper: Experimental transmission spectra with changing pump powers, with values of \SI{1}{\micro\watt} (red), \SI{5}{\micro\watt}, \SI{10}{\micro\watt},  \SI{50}{\micro\watt}, \SI{100}{\micro\watt}, \SI{500}{\micro\watt}, \SI{1000}{\micro\watt}, \SI{7000}{\micro\watt} (purple). Lower: Modelled transmission spectra with pump rabi frequencies of \SI{1}{\mega\Hz} (red), \SI{2}{\mega\Hz}, \SI{3}{\mega\Hz}, \SI{5}{\mega\Hz}, \SI{10}{\mega\Hz}, \SI{20}{\mega\Hz}, \SI{30}{\mega\Hz}, \SI{50}{\mega\Hz}, \SI{100}{\mega\Hz}, \SI{300}{\mega\Hz}, \SI{500}{\mega\Hz} (purple).}
% \label{fig:change-power-780-pump}
% \end{figure}

\section{Conclusions and Outlook}

In conclusion, we have observed a clean, narrow EIT feature in a V-system and the concomitant enhanced absorption. We see that the EIT feature has contributions from a coherent process, and an incoherent optical pumping process. The incoherent contribution occurs because of the allowed decay from the excited states to both ground states, and is the cause of the enhanced absorption feature. Our theoretical model captures all of the relevant processes, and gives insight into the role of coherence in explaining the observed narrow spectral features. The theoretical treatment is greatly simplified because the experiment was conducted in the hyperfine Paschen--Back regime, leading  to distinct non-overlapping resonances.

In this work the Doppler mismatch is small, however the clean system presented here would easily allow investigation of the effect of large mismatches, for example the 5S-5P 5S-6P V-system in rubidium \cite{PhysRevA.57.1323,VDOVIC2007407},  and could be the subject of further study. 
Another interesting aspect of this work is the sensitivity of the V-EIT spectra to collisional linewidths. Indeed, the study of isolated resonances in the hyperfine Paschen-Back regime could provide a more sensitive spectral measurement of alkali metal--noble gas collisions, and will be the subject of future work.

% \section{Supplemental Material}

% Consult the Author Guidelines for Supplementary Materials in OSA Journals for details on accepted types of materials and instructions on how to cite them.
% All materials must be associated with a figure, table, or equation or be referenced in the results section of the manuscript.
% (1) 2D and 3D image files and video must be labeled “Visualisation,” not “Movie,” “Video,” “Figure,” etc.
% (2) Machine-readable data (for example, csv files) must be labeled  “Data File.”  Number data files and visualisations consecutively, e.g., “Visualisation 1, Visualisation 2….”
% (3) Large datasets or code files must be placed in an open, archival database.  Such items should be mentioned in the text as either “Dataset” or “Code,” as appropriate, and also be cited in the references list.  For example, “see Dataset 1 (Ref. [1]) and Code 1 (Ref [2]).” Here are examples of the references:

\noindent\textbf{Funding.} EPSRC (EP/R002061/1); Durham University.

\noindent\textbf{Acknowledgements.} The authors thank Daniel Whiting for help with the model code, and Renju Mathew, Danielle Pizzey, Lina Marieth Hoyos-Campo and Tom Cutler for helpful discussions. The data presented in this paper are available from DRO, https:// doi:10.15128/r2t148fh16h.

\noindent\textbf{Disclosures.} The authors declare no conflicts of interest.

% Bibliography
\bibliographystyle{unsrt}
\bibliography{references}

\begin{thebibliography}{10}

\bibitem{ARIMONDO1996257}
E.~Arimondo.
\newblock V. {C}oherent population trapping in laser spectroscopy.
\newblock volume~35 of {\em Progress in Optics}, pages 257--354. Elsevier,
  1996.

\bibitem{Fleischhauer:magnetometry}
M.~Fleischhauer, A.~B. Matsko, and M.~O. Scully.
\newblock Quantum limit of optical magnetometry in the presence of ac {S}tark
  shifts.
\newblock {\em Phys. Rev. A}, 62:013808, Jun 2000.

\bibitem{Budker2007:magnetometry}
Dmitry Budker and Michael Romalis.
\newblock Optical magnetometry.
\newblock {\em Nature Physics}, 3(4):227--234, Apr 2007.

\bibitem{Yudin:magnetometry}
V.~I. Yudin, A.~V. Taichenachev, Y.~O. Dudin, V.~L. Velichansky, A.~S. Zibrov,
  and S.~A. Zibrov.
\newblock Vector magnetometry based on electromagnetically induced transparency
  in linearly polarized light.
\newblock {\em Phys. Rev. A}, 82:033807, Sep 2010.

\bibitem{Hau:slowlight}
L.~Hau, S.~Harris, and Cyrus~H. Dutton, Z.and~Behroozi.
\newblock Light speed reduction to 17 metres per second in an ultracold atomic
  gas.
\newblock {\em Nature}, 397:594–598, 1999.

\bibitem{Das_2018}
Arpita Das, Bankim~Chandra Das, Dipankar Bhattacharyya, Shrabana Chakrabarti,
  and Sankar De.
\newblock Polarization rotation with electromagnetically induced transparency
  in a {V}-type configuration of {R}b {D}1 and {D}2 transitions.
\newblock {\em Journal of Physics B: Atomic, Molecular and Optical Physics},
  51(17):175502, aug 2018.

\bibitem{Beausoleil:quantuminformation}
R.~G. Beausoleil, W.~J. Munro, D.~A. Rodrigues, and T.~P. Spiller.
\newblock Applications of electromagnetically induced transparency to quantum
  information processing.
\newblock {\em Journal of Modern Optics}, 51(16-18):2441--2448, 2004.

\bibitem{Hammerer:quantuminformation}
Klemens Hammerer, Anders~S. S\o{}rensen, and Eugene~S. Polzik.
\newblock Quantum interface between light and atomic ensembles.
\newblock {\em Rev. Mod. Phys.}, 82:1041--1093, Apr 2010.

\bibitem{Santra:atomicclock}
Robin Santra, Ennio Arimondo, Tetsuya Ido, Chris~H. Greene, and Jun Ye.
\newblock High-accuracy optical clock via three-level coherence in neutral
  bosonic $^{88}\mathrm{Sr}$.
\newblock {\em Phys. Rev. Lett.}, 94:173002, May 2005.

\bibitem{FleischhauerReview}
Michael Fleischhauer, Atac Imamoglu, and Jonathan~P. Marangos.
\newblock {Electromagnetically induced transparency: Optics in coherent media}.
\newblock {\em Rev. Mod. Phys.}, 77:633--673, Jul 2005.

\bibitem{KHAN20164100}
Sumanta Khan, Vineet Bharti, and Vasant Natarajan.
\newblock Role of dressed-state interference in electromagnetically induced
  transparency.
\newblock {\em Physics Letters A}, 380(48):4100--4104, 2016.

\bibitem{DEY20152711}
S.~Dey, S.~Mitra, P.N. Ghosh, and B.~Ray.
\newblock {EIT line shape in an open and partially closed multilevel V-type
  system}.
\newblock {\em Optik}, 126(20):2711--2717, 2015.

\bibitem{Hazra_2020}
Rohit Hazra and Md~Mabud Hossain.
\newblock Study of multi-window electromagnetically induced transparency
  ({EIT}) and related dispersive signals in {V}-type systems in the {Z}eeman
  sublevels of hyperfine states of 87{R}b-{D}2 line.
\newblock {\em Journal of Physics B: Atomic, Molecular and Optical Physics},
  53(23):235401, oct 2020.

\bibitem{Hoshina:14}
Yoshitaka Hoshina, Nobuhito Hayashi, Kosuke Tsubota, Ichiro Yoshida, Kotaro
  Shijo, Ryuta Sugizono, and Masaharu Mitsunaga.
\newblock {Electromagnetically induced transparency in a V-type multilevel
  system of Na vapor}.
\newblock {\em J. Opt. Soc. Am. B}, 31(8):1808--1813, Aug 2014.

\bibitem{PhysRevA.57.1323}
J.~R. Boon, E.~Zekou, D.~J. Fulton, and M.~H. Dunn.
\newblock {Experimental observation of a coherently induced transparency on a
  blue probe in a Doppler-broadened mismatched V-type system}.
\newblock {\em Phys. Rev. A}, 57:1323--1328, Feb 1998.

\bibitem{PhysRevA.83.063419}
A.~Lazoudis, T.~Kirova, E.~H. Ahmed, P.~Qi, J.~Huennekens, and A.~M. Lyyra.
\newblock {Electromagnetically induced transparency in an open V-type molecular
  system}.
\newblock {\em Phys. Rev. A}, 83:063419, Jun 2011.

\bibitem{PhysRevA.59.4675}
J.~R. Boon, E.~Zekou, D.~McGloin, and M.~H. Dunn.
\newblock Comparison of wavelength dependence in cascade-,
  \ensuremath{\Lambda}-, and {V}ee-type schemes for electromagnetically induced
  transparency.
\newblock {\em Phys. Rev. A}, 59:4675--4684, Jun 1999.

\bibitem{PhysRevA.87.043813}
Chengjie Zhu, Chaohua Tan, and Guoxiang Huang.
\newblock {Crossover from electromagnetically induced transparency to
  Autler-Townes splitting in open V-type molecular systems}.
\newblock {\em Phys. Rev. A}, 87:043813, Apr 2013.

\bibitem{McGloin_2003}
D~McGloin.
\newblock {Coherent effects in a driven Vee scheme}.
\newblock {\em Journal of Physics B: Atomic, Molecular and Optical Physics},
  36(13):2861--2871, jun 2003.

\bibitem{PhysRevA.71.053806}
Ying Wu and Xiaoxue Yang.
\newblock Electromagnetically induced transparency in {V}-,
  $\ensuremath{\Lambda}$-, and cascade-type schemes beyond steady-state
  analysis.
\newblock {\em Phys. Rev. A}, 71:053806, May 2005.

\bibitem{ZHAO2002341}
Jianming Zhao, Lirong Wang, Liantuan Xiao, Yanting Zhao, Wangbao Yin, and
  Suotang Jia.
\newblock {Experimental measurement of absorption and dispersion in V-type
  cesium atom}.
\newblock {\em Optics Communications}, 206(4):341--345, 2002.

\bibitem{PhysRevA.92.063810}
S.~Scotto, D.~Ciampini, C.~Rizzo, and E.~Arimondo.
\newblock {Four-level $\mathsf{N}$-scheme crossover resonances in Rb saturation
  spectroscopy in magnetic fields}.
\newblock {\em Phys. Rev. A}, 92:063810, Dec 2015.

\bibitem{CHA2014175}
Eun~Hyun Cha, Taek Jeong, and Heung-Ryoul Noh.
\newblock Two-color polarization spectroscopy in {V}-type configuration in
  rubidium.
\newblock {\em Optics Communications}, 326:175--179, 2014.

\bibitem{Das:21}
Arpita Das, Bankim~Chandra Das, Dipankar Bhattacharyya, and Sankar De.
\newblock Effects of probe ellipticity and longitudinal magnetic field on the
  polarization rotation in a coherently prepared atomic medium.
\newblock {\em OSA Continuum}, 4(1):105--120, Jan 2021.

\bibitem{VDOVIC2007407}
Silvije Vdović, Ticijana Ban, Damir Aumiler, and Goran Pichler.
\newblock {EIT} at 5$^2${S}$_{1/2}$→6$^2${P}$_{3/2}$ transition in a
  mismatched {V}-type rubidium system.
\newblock {\em Optics Communications}, 272(2):407--413, 2007.

\bibitem{Kang:17}
Hyun-Jong Kang and Heung-Ryoul Noh.
\newblock {Coherence effects in electromagnetically induced transparency in
  V-type systems of 87Rb}.
\newblock {\em Opt. Express}, 25(18):21762--21774, Sep 2017.

\bibitem{PhysRevA.52.2302}
David~J. Fulton, Sara Shepherd, Richard~R. Moseley, Bruce~D. Sinclair, and
  Malcolm~H. Dunn.
\newblock {Continuous-wave electromagnetically induced transparency: A
  comparison of V, $\Lambda$, and cascade systems}.
\newblock {\em Phys. Rev. A}, 52:2302--2311, Sep 1995.

\bibitem{PhysRevA.84.063410}
B.~A. Olsen, B.~Patton, Y.-Y. Jau, and W.~Happer.
\newblock {Optical pumping and spectroscopy of Cs vapor at high magnetic
  field}.
\newblock {\em Phys. Rev. A}, 84:063410, Dec 2011.

\bibitem{Weller_2012}
Lee Weller, Kathrin~S Kleinbach, Mark~A Zentile, Svenja Knappe, Charles~S
  Adams, and Ifan~G Hughes.
\newblock {Absolute absorption and dispersion of a rubidium vapour in the
  hyperfine Paschen-Back regime}.
\newblock {\em Journal of Physics B: Atomic, Molecular and Optical Physics},
  45(21):215005, oct 2012.

\bibitem{Zentile_2014}
Mark~A Zentile, Rebecca Andrews, Lee Weller, Svenja Knappe, Charles~S Adams,
  and Ifan~G Hughes.
\newblock {The hyperfine Paschen--Back Faraday effect}.
\newblock {\em Journal of Physics B: Atomic, Molecular and Optical Physics},
  47(7):075005, mar 2014.

\bibitem{Ponciano_Ojeda_2020}
Francisco~S Ponciano-Ojeda, Fraser~D Logue, and Ifan~G Hughes.
\newblock {Absorption spectroscopy and Stokes polarimetry in a 87Rb vapour in
  the Voigt geometry with a 1.5 T external magnetic field}.
\newblock {\em Journal of Physics B: Atomic, Molecular and Optical Physics},
  54(1):015401, dec 2020.

\bibitem{Sargsyan:14}
A.~Sargsyan, G.~Hakhumyan, C.~Leroy, Y.~Pashayan-Leroy, A.~Papoyan,
  D.~Sarkisyan, and M.~Auzinsh.
\newblock Hyperfine {P}aschen-{B}ack regime in alkali metal atoms: consistency
  of two theoretical considerations and experiment.
\newblock {\em J. Opt. Soc. Am. B}, 31(5):1046--1053, May 2014.

\bibitem{Sargsyan:17a}
Armen Sargsyan, Emmanuel Klinger, Grant Hakhumyan, Ara Tonoyan, Aram Papoyan,
  Claude Leroy, and David Sarkisyan.
\newblock {Decoupling of hyperfine structure of Cs D1 line in strong magnetic
  field studied by selective reflection from a nanocell}.
\newblock {\em J. Opt. Soc. Am. B}, 34(4):776--784, Apr 2017.

\bibitem{PhysRevA.95.061804}
L.~Ma, D.~A. Anderson, and G.~Raithel.
\newblock {Paschen-Back effects and Rydberg-state diamagnetism in vapor-cell
  electromagnetically induced transparency}.
\newblock {\em Phys. Rev. A}, 95:061804, Jun 2017.

\bibitem{doi:10.1063/1.4993760}
S.~George, N.~Bruyant, J.~Béard, S.~Scotto, E.~Arimondo, R.~Battesti,
  D.~Ciampini, and C.~Rizzo.
\newblock Pulsed high magnetic field measurement with a rubidium vapor sensor.
\newblock {\em Review of Scientific Instruments}, 88(7):073102, 2017.

\bibitem{WhitingEIT}
Daniel~J. Whiting, James Keaveney, Charles~S. Adams, and Ifan~G. Hughes.
\newblock {Direct measurement of excited-state dipole matrix elements using
  electromagnetically induced transparency in the hyperfine Paschen-Back
  regime}.
\newblock {\em Phys. Rev. A}, 93:043854, Apr 2016.

\bibitem{WhitingEIA}
Daniel~J. Whiting, Erwan Bimbard, James Keaveney, Mark~A. Zentile, Charles~S.
  Adams, and Ifan~G. Hughes.
\newblock Electromagnetically induced absorption in a nondegenerate three-level
  ladder system.
\newblock {\em Opt. Lett.}, 40(18):4289--4292, Sep 2015.

\bibitem{Whiting:FWM}
D.~J. Whiting, Renju~S. Mathew, J.~Keaveney, C.~S. Adams, and I.~G. Hughes.
\newblock {Four-wave mixing in a non-degenerate four-level diamond
  configuration in the hyperfine Paschen–Back regime}.
\newblock {\em Journal of Modern Optics}, 65(5-6):713--722, 2018.

\bibitem{WhitingSinglephotons}
D.~J. Whiting, N.~\ifmmode \check{S}\else
  \v{S}\fi{}ibali\ifmmode~\acute{c}\else \'{c}\fi{}, J.~Keaveney, C.~S. Adams,
  and I.~G. Hughes.
\newblock Single-photon interference due to motion in an atomic collective
  excitation.
\newblock {\em Phys. Rev. Lett.}, 118:253601, Jun 2017.

\bibitem{f2f}
C.~S. Adams and I.~G. Hughes.
\newblock {\em {Optics f2f - From Fourier to Fresnel}}.
\newblock Oxford University Press, Oxford, UK, 2019.

\bibitem{PhysRevA.73.062509}
M.~L. Harris, C.~S. Adams, S.~L. Cornish, I.~C. McLeod, E.~Tarleton, and I.~G.
  Hughes.
\newblock Polarization spectroscopy in rubidium and cesium.
\newblock {\em Phys. Rev. A}, 73:062509, Jun 2006.

\bibitem{doi:10.1143/JPSJ.78.084302}
Seo Ro~Shin and Heung-Ryoul Noh.
\newblock Calculation and measurement of absolute transmission in rubidium.
\newblock {\em Journal of the Physical Society of Japan}, 78(8):084302, 2009.

\bibitem{Sagle_1996}
J~Sagle, R~K Namiotka, and J~Huennekens.
\newblock Measurement and modelling of intensity dependent absorption and
  transit relaxation on the cesium line.
\newblock {\em Journal of Physics B: Atomic, Molecular and Optical Physics},
  29(12):2629--2643, jun 1996.

\bibitem{doi:10.1080/09500340.2017.1328749}
Ifan~G. Hughes.
\newblock {Velocity selection in a Doppler-broadened ensemble of atoms
  interacting with a monochromatic laser beam}.
\newblock {\em Journal of Modern Optics}, 65(5-6):640--647, 2018.

\bibitem{Zentile2015b}
Mark~A. Zentile, James Keaveney, Lee Weller, Daniel~J. Whiting, Charles~S.
  Adams, and Ifan~G. Hughes.
\newblock {ElecSus: A program to calculate the electric susceptibility of an
  atomic ensemble}.
\newblock {\em Comput. Phys. Commun.}, 189:162--174, apr 2015.

\end{thebibliography}
%\bibliography{ref_test}

% Full bibliography added automatically for Optics Letters submissions; the following line will simply be ignored if submitting to other journals.
% Note that this extra page will not count against page length
%\bibliographyfullrefs{references}
%\bibliographyfullrefs{ref_test}

\end{document}